Inducing superconductivity in quantum anomalous Hall regime

Yu Huang[1,2,3], Yu Fu[1,2,3], Peng Zhang[4], Kang L. Wang[4], Qing Lin He[1,2,3*]

[1]International Center for Quantum Materials, School of Physics, Peking University, Beijing 100871, China

[2]Collaborative Innovation Center of Quantum Matter, Beijing 100871, China

[3]Interdisciplinary Institute of Light-Element Quantum Materials and Research Center for Light-Element Advanced Materials, Peking University, Beijing 100871, China

[4]Department of Electrical and Computer Engineering, Department of Physics and Astronomy, and Department of Materials Science and Engineering, University of California, Los Angeles, CA 90095, USA

[*]Email: qlhe@pku.edu.cn

**Abstract**

Interfacing the quantum anomalous Hall insulator with a conventional superconductor is known to be a promising manner for realizing a topological superconductor, which has been continuously pursued for years. Such a proximity route depends to a great extent on the control of the delicate interfacial coupling of the two constituents. However, a recent experiment reported the failure to reproduce such a topological superconductor, which is ascribed to the negligence of the electrical short by the superconductor in the theoretical proposal. Here, we reproduce this topological superconductor with attention to the interface control. The resulted conductance matrix under a wide magnetic field range agrees with the fingerprint of this topological superconductor. This allows us to develop a phase diagram that unveils three regions parameterized by various coupling limits, which not only supports the feasibility to fabricate the topological superconductor by proximity but also fully explains the origin of the previous debate. The present work provides a comprehensible guide on fabricating the topological superconductor.

## 1. Introduction

Superconductors (SCs) with nontrivial topology, dubbed topological SCs, are predicted to accommodate Majorana bound states that exhibit non-Abelian statistics and thus provide potential to construct topological qubits for fault-tolerant topological quantum computation [1-4]. The realization of topological SCs has been experimentally pursued using various platforms, yet remains a long-standing challenge. One of the attractive routes refers to the quantum anomalous Hall (QAH) insulator that acquires superconductivity via proximity effect [2,5]. In this system, time-reversal symmetry is broken by the mass term $m$, which acts as a knob to control the chiral fermion and Majorana edge states classified by Chern number $\mathcal{C}$ and Bogoliubov-de Gennes Chern number $\mathcal{N}$, respectively. Considerable efforts have been devoted to the fabrication and detection of this topological SC. Spectroscopic evidences for the presences of chiral and bound Majorana states had been discovered in QAH/SC heterostructures [6,7], while the crossed Andreev reflection was observed in a similar heterostructure [8]. These examples provide evidences of proximity-induced superconductivity in QAHs. The primitive constituent of a topological qubit, *i.e.* the QAH/SC/QAH junction, was proposed [9,10] that could manifest distinct fingerprints of Majorana states. This inspired further theoretical proposals [11-17] towards constructing qubits. Unfortunately, a recent experiment [18] reported the failure to reproduce these fingerprints using the proposed junction. Instead of following the theoretical evolution described by $\mathcal{C} = \pm 1 \rightarrow 0$ and $\mathcal{N} = \pm 2 \rightarrow \pm 1 \rightarrow 0$, the junction conductance was found to be always half-quantized at $\frac{e^2}{2h}$, which was ascribed to the electrical short by SC. This result casts doubts about the feasibility of inducing superconductivity to QAH via proximity using this junction. However, it is well known that the effectiveness of proximity is highly sensitive to the interfacial coupling. While for the scenario that two constituents directly contact with each other, the proximity effectiveness seems to be maximal, but there may be an electrical short. Conversely, the interface will contain a nonsuperconducting dead layer that exists at the interface between the non-SC substrate and SC film [19-22]. This could be caused by the lowering of electron density and/or weakening of phonon coupling due to interdiffusions, strains, defects, grain boundaries, off-stoichiometries, or other growth-related factors at the interface [23-27]. Hence, the proximity effectiveness may not be maximized but rather provide additional conductive channel across the QAH surface, which results in the observed electrical short. To check this, a methodology that can quantitatively control the proximity is needed.

In this work, the proximity control is achieved by inserting $AlO_x$ layers of various thicknesses ($t$) at the QAH/SC interface or a native surface oxide on QAH, via which we realize the proximity-induced superconductivity in QAH. By probing the conductance matrix in a wide magnetic field range, we found three types of responses categorized by two $t$-thresholds. Between the strong coupling and decoupling limits, the proximity effect takes the leading role, in which the junction responses consistently with the theory. We unveil a phase diagram that could fully conform the present results with those from Ref. [18], which demonstrates that, in the QAH/SC/QAH junction, not only the electrical short can be suppressed but also the superconductivity can be induced to QAH.

## 2. Experimental Methods

### 2.1. Films growths and junctions fabrications

All the QAHs consist of 6-nm-thick Cr:$(Bi,Sb)_2Te_3$ films epitaxially grown on semi-insulating GaAs(111)B substrates using a molecular beam epitaxy technique. The QAH film, whose native surface oxide shows the effectiveness in interface control, was stored in a sample box for about 20 days before the magnetoelectrical measurement was performed. Bars of QAH (width: 350 $\mu$m) were patterned by a set of stencil masks using a reactive ion etching technique. After this, continuous $AlO_x$ layers of various thicknesses ($t$ = 0, 0.5, 0.7, 0.8, 0.9, 1.1, 1.3, 1.5, 2, and 3 nm) were sputtered (rate: 0.03 Å/s) onto the entire samples surfaces. Then another set of stencil masks was stacked on the samples surfaces before the sputterings of Nb (rate: 0.52 Å/s; thickness: 100 nm; width: 500 $\mu$m), followed by sputterings of Al (thickness: 3 nm; rate: 0.50 Å/s) to avoid the oxidation of Nb. Contacts were made by indium and Ag wires.

### 2.2. Magnetoelectrical measurements

The magnetoelectrical transport measurements were performed in a dilution refrigerator at 20 mK with the magnetic fields applied perpendicular to the devices surfaces. The junction fabricated using the surface-oxidized QAH was measured in another dilution refrigerator at 50 mK. The AC current was applied across a 10-MΩ reference resistor by a small voltage, while voltages across the devices were measured by the lock-in amplifier (SR865A). Another lock-in amplifier (SR865A) was used to measure the current through the device.

## 3. Results and discussion

### 3.1. Phase diagram

By modulating the interfacial coupling, the junction conductance shows evolutions among three trivial and nontrivial regions. To quantitatively control the coupling, $AlO_x$ insertion layers of various nominal thicknesses ($t$=0, 0.5, 0.7, 0.8, 0.9, 1.1, 1.3, 1.5, 2, and 3 nm) were deposited at the QAH/SC interface. The QAH effect in these QAH/SC/QAH junctions are still maintained, as supported by the three-terminal resistances measurements in Fig. S1 of Supplementary Material. Inset of Fig. 1(a) displays the measurement configuration of the conductance $\sigma_{12}$ across the junction. When QAH and SC directly contact to each other (corresponding to $t = 0$), $\sigma_{12}$ keeps ~$0.45\frac{e^2}{h}$ under a perpendicular magnetic field $B_z$ of ±0.5 T in Fig. 1(a), similar to those in Ref. [18]. The slight deviation from $\frac{e^2}{2h}$ probably originates in the minor degradation of QAH after being partially capped by Nb. This is supported by the result that, after inserting an AlOx layer of $t$=0.5 nm that protects the QAH surface, $\sigma_{12}$ shows the exact half-quantization. It is clear that both scenarios of $t$=0 and 0.5 nm are resulted from electrical shorts by the Nb bar traversing QAH as observed in Ref. [18]. However, in the following scenarios that when such a direct contact is avoided by modulating the interfacial coupling, *i.e.* varying 0.5 nm<$t$<1.5 nm, the electrical short is suppressed whereas the proximity effect persists. In Fig. 1(b), junctions with these $t$ exhibit various conductance increments to $0.71\frac{e^2}{h}<\sigma_{12}<0.97\frac{e^2}{h}$ in QAH states, in sharp contrast to the half-quantizations in Fig. 1(a) and Ref. [18]. We notice that some discontinuity arises among these $\sigma_{12}$ along with the gradual change of $t$. This is because, on one hand, as these junctions fabrications are not fully *in-situ*, uncontrollable variants such as oxidation and surface contamination may exist at the interfaces of AlOx/QAH and/or Nb/AlOx within these junctions. On the other hand, owing to these variants and consequently the roughness of the AlOx layer, the nominal $t$ cannot reflect the effective thickness of AlOx, $t_{\text{eff}}$, which determines the tunneling probability of Cooper pairs into QAH. When $t$>1.5 nm, both constituents fully decouple from each other. We observed consistent behaviors of $\sigma_{12}$~$\frac{e^2}{h}$ in QAH states in Fig. 1(c), which solely come from the quantum transport of QAH without any superconductivity since this transport remains intact at high-$B_z$ where the superconductivity in Nb is fully eliminated (to be addressed below). Such a high-$B_z$ response dramatically differs from those when 0.5 nm<$t$<1.5 nm as presented later, which demonstrates their different physical origins. Therefore, the above evolution can be summarized as a phase diagram parameterized by the $t$-dependent $\sigma_{12}$ under 0.5 T in Fig. 1(d). One can find that $\sigma_{12}$ evolves as a function of the interfacial coupling modulated by $t$, which results in three regions: Region I is trivially dominated by the electrical short that does not allow the distinction from proximity effect; Region III is another extreme that the AlO$_x$ layer is sufficiently thick and proximity effect is eliminated; Region II is intermediate that not only suppresses the electrical short but also allows the Cooper pairs to tunnel from SC into QAH, leading to a modulated proximity effect. As discussed above, results in Ref. [18] should locate within Region I [hollow square in Fig. 1(d)].

### 3.2. Signature of Majoranas

In Region II, signature of Majorana states is illustrated. When $m$ of QAH is modulated by $B_z$, QAH evolves as $\mathcal{C}=\pm 1 \rightarrow 0$ with the Hall conductance varied by $\sigma_{xy}=\pm\frac{e^2}{h}\rightarrow 0$, while the corresponding topological SC evolves as $\mathcal{N}=\pm 2\rightarrow 0$ with $\sigma_{12}=\frac{e^2}{h}\rightarrow 0$, during which single Majorana states of $\mathcal{N}=\pm 1$ appear with $\sigma_{12}=\frac{e^2}{2h}$. Signature of this evolution is exemplified in the junction with $t = 1.3$ nm in Fig. 2(a-b) and $t = 0.9$, 1.1 nm in Figs. S2-3 of Supplementary Material. When $\sigma_{xy}$ is about to deviate from $\pm\frac{e^2}{h}$ towards 0 (light red regions and red arrows) and just enter $\pm\frac{e^2}{h}$ from 0 (light blue regions and blue arrows), a pair of kinks of $\sigma_{12}\sim 0.57\frac{e^2}{h}$ and $\sim 0.59\frac{e^2}{h}$ arises, respectively, which is close to $\frac{e^2}{2h}$ in theory. In spite of the fact that these kinks are very narrow and inconspicuous, they could be repeatedly captured within the same $B_z$-range during multiple $B_z$-sweeps (also see repeatability tests using junctions with $t$=0.9 and 1.1 nm in Figs. S2-4 of Supplementary Material) and after thermal cycling (Fig. S4 of Supplementary Material for repeatability tests after increasing to 35 K, a temperature above the Curie temperature ~30 K of QAH [28]). These observations agree with the theory [10,12] that states of $\mathcal{N}=\pm 1$ only appear when about to leave from or enter QAH states with $\sigma_{xy}$ quantized. The values of these kinks seem to be always slightly higher than $\frac{e^2}{2h}$, which indicates the existence of additional conductive channels probably from percolations [29-32] and/or Nb. States of $\mathcal{N}=\pm 2$ are signified by $\sigma_{12} \sim 0.74\frac{e^2}{h}$ plateaus, which deviate from the ideal quantization $\frac{e^2}{h}$ probably because of the inhomogeneity of the induced superconductivity and/or the parallel channels from Nb. The above $\sigma_{12}$ are essentially similar to those in a previous study [33,34] except the full quantizations values.

### 3.3. Conductance across interface

The proximity effect is also evidenced by the conductance across the QAH/SC interface, $\sigma_{13}$, whose measurement configuration is shown in the inset of Fig. 3. It has been shown above that the Nb bar will function as a normal electrode and its superconductivity plays no role for junctions with $t=0$ and 0.5 nm in Region I. In this way, $\sigma_{13}$ is equivalent to the superposition of magnetoelectrical and Hall conductances of a QAH, which should quantize at $\frac{e^2}{h}$. This is revealed by Fig. 3(a), in which $\sigma_{13}$ of the two junctions quantize at ~$0.87\frac{e^2}{h}$ and $\frac{e^2}{h}$ respectively, consistent with our conclusion regarding Region I and similar to those in Ref. [18]. In contrast to this trivial behavior, $\sigma_{13}$ of junctions in Region II, as exemplified by the ones with $t=0.7$ and 1.3 nm in Fig. 3(b), exhibit four peaks around $B_z$-ranges of $\mathcal{N}=\pm 1$ states. As shown in Supplementary Text I of Supplementary Material, in the limit of negligible contact resistance and due to a large amount of conductive channels within SC as a probe, $\sigma_{13}$ should be close to $\frac{e^2}{h}$ around $\mathcal{N}=\pm 1$ states, but almost vanish otherwise. This trend is essentially captured by the resulted two $\sigma_{13}$, saving that the peaks values are still far from quantization, partly because of the contact resistance in the three-probe method. Such a behavior has never been captured in Ref. [18]. In the junction with $t=2$ nm from Region III where QAH and SC decouple, the transmission across the interface is prohibited and $\sigma_{13}$ becomes plain. Consistent to $\sigma_{12}$, the present $\sigma_{13}$ is also modulated by $t$ and show distinct responses in the three regions of the phase diagram, which demonstrates the effectiveness of interface control.

### 3.4. High-field responses

Results measured under strong $B_z$ that destroy the superconductivity also support our findings. When $B_z$ is higher than the upper critical field of Nb, *i.e.* $B_c=3.0\sim3.7$ T at 20 mK (Fig. S5 of Supplementary Material), the superconductivity is eliminated. The Nb bar then acts as a normal electrode that connects the neighboring QAH bars in series, which results in a 50% decrease of $\sigma_{12}$ from $\frac{e^2}{h}$ when $B_z < B_c$ to $\frac{e^2}{2h}$ when above. Such a $B_c$-dependent behavior could be regarded as another fingerprint of a superconducting QAH because $\sigma_{12}$ in trivial scenarios always show independences on high-$B_z$. That is, $\sigma_{12}$ keeps at $\frac{e^2}{2h}$ if QAH is electrically shorted, or stays at $\frac{e^2}{h}$ when decouples. To check this, our junctions are subjected to sweepings under high-$B_z$ in Fig. 4. As opposed to the constant $\sigma_{12}\sim\frac{e^2}{2h}$ in Ref. [18], $\sigma_{12}$ of our junctions in Region II decrease by 14~44% at 10 T with respect to those at 1 T. Consider that the resistances of Nb bars keep almost constant (variations <1 Ω) when both $B_z < B_c$ and $B_z > B_c$ (Fig. S5 of Supplementary Material), such substantial decreases in $\sigma_{12}$ should be dominated by the suppressions against the induced superconductivities in QAHs, instead of the parallel channel from Nb. These induced superconductivities originate in the proximity effect from Nb since these decreases become prominent [labeled by arrows in Fig. 4(a)] around $B_c$, which marks the disappearance of superconductivity from Nb. However, again, these decreases vary inconsistently with the gradual change of $t$ owing to the uncontrollable interfacial variants introduced by *ex-situ* fabrications discussed above, whereas their $t_\text{eff}$ remain to be further explored. The continuous decreases of $\sigma_{12}$ when above $B_c$ probably relate to the field-dependent tunneling across the interface, which could cause a result similar with the electrical short. When Nb strongly couples to and decouples from QAH like those in Regions I and III, respectively, $\sigma_{12}$ simply stay constant around $\frac{e^2}{2h}$ and $\frac{e^2}{h}$ at high-$B_z$ in Fig. 4(b). Such distinct high-$B_c$ responses from junctions in Regions II *vs.* I and III underline the induced superconductivity in QAH. These high-$B_z$ responses are also plotted in the phase diagram in Fig. 1(d), which show consistency with the low-$B_z$ one in identifying the phases. The induced superconductivity is demonstrated to locate within Region II by the high-$B_z$-induced decrease in $\sigma_{12}$ as highlighted by the shaded region.

## 4. Conclusion

Besides inserting the AlO$_x$ layer to modulate the interfacial coupling, the native surface oxide of some QAH could play a similar role. Although $t_\text{eff}$ of the surface oxide is unquantifiable either, the junction fabricated using such a QAH (still shows quantized Hall conductivity and vanishing longitudinal conductivity) without any AlO$_x$ insertion can also show signature of Majorana states, as evidenced by $\sigma_{xy}$ and $\sigma_{12}$ in Fig. S6 of Supplementary Material. The pair of half-quantizations in $\sigma_{12}$ is 0.53~0.54$\frac{e^2}{h}$ while the full quantization is 0.71~0.76$\frac{e^2}{h}$, similar to those with appropriate AlO$_x$ insertions. Under high-$B_z$, $\sigma_{12}$ exhibits a prompt decrease when $B_z < B_c$ in prior to a saturation to ~0.6$\frac{e^2}{h}$ when above (Fig. S7 of Supplementary Material), similar to the previous study [33,34].

Better quantizations may be achieved in our junctions of Region II if the uniformity and thickness of the AlO$_x$ layer and/or native surface oxide are further optimized, so that the induced superconductivity becomes more homogeneous whereas parallel conductance from Nb is fully excluded. An all-*in-situ* fabrication process could fully address these interfacial uncertainties and may achieve the expected quantizations. The present work provides a broad view from the phase diagram and reveals that the theoretical predictions [9,10] still hold, which lays out the experimental framework in constructing the topological SC and qubit. We notice that another experimental study [8] reports the observation of the crossed Andreev reflection also arising from the induced superconductivity in QAH by Nb, whose results were also shown to be different from Ref. [18], which led to the conclusion therein of no evidence for any Andreev process occurred in Ref. [18].

**Acknowledgments**

This research is supported by the National Key RD Program of China (Grants No. 2020YFA0308900) and the Strategic Priority Research Program of Chinese Academy of Sciences (Grant No. XDB28000000). The research in UCLA is supported in part by the U.S. Army Research Office MURI program (Grants No. W911NF-16-1-0472 and W911NF-20-2-0166), and NSF grants (No. 1611570, 1936383, and 2040737).

**Conflict of interest**

The authors have no conflicts to disclose.

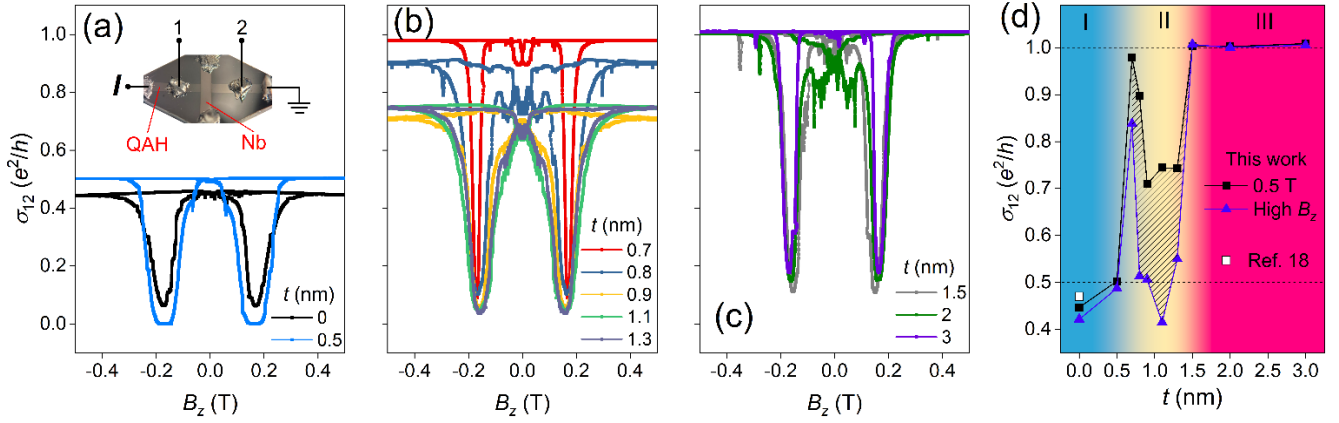

FIG. 1. (a-c) Conductances $\sigma_{12}$ across the QAH/SC/QAH junctions with various $AlO_x$ layers thicknesses $t$ as functions of perpendicular magnetic fields $B_z$. Inset in (a) shows the measurement configuration. QAH, quantum anomalous Hall. (d) Summary of $\sigma_{12}$ under 0.5 T and high-$B_z$ (5 T for $t$=0, 0.5 nm; 7 T for $t$=1.5 nm; 8 T for $t$=2 nm; 10 T for others; see Fig. 4) as functions of $t$. Three regions are identified based on the different responses of $\sigma_{12}$ to the low- and high-$B_z$, while the high-$B_z$-induced decrease of $\sigma_{12}$ (shaded area) essentially locates within Region II.

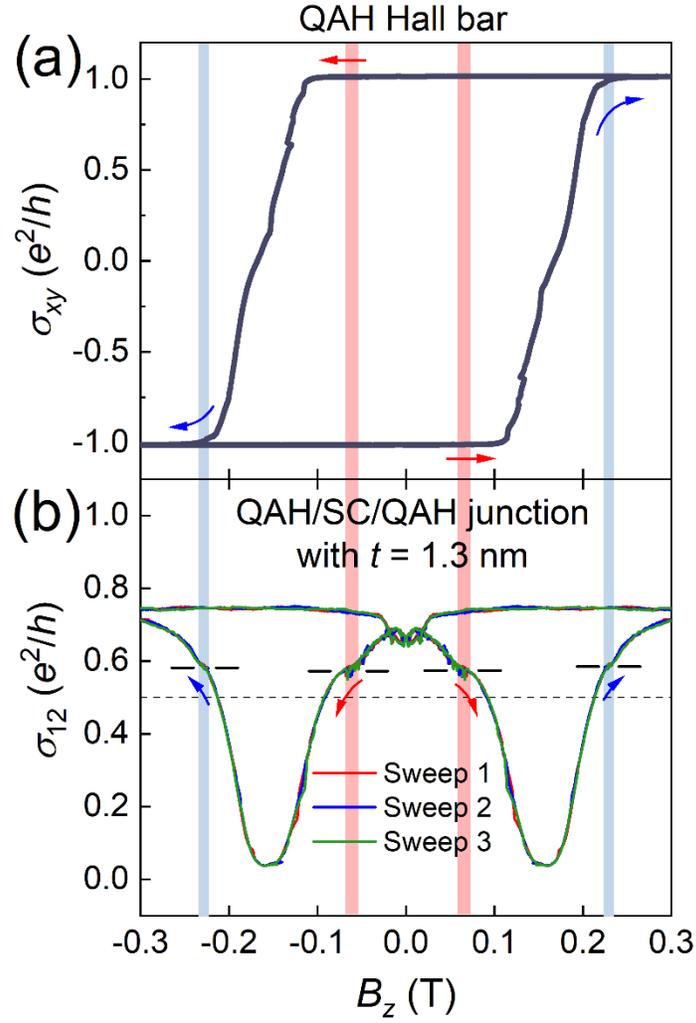

FIG. 2. (a) Hall conductance $\sigma_{xy}$ of a QAH, and (b) the conductances $\sigma_{12}$ across the QAH/SC/QAH junction with the AlO$_x$ thickness $t$ = 1.3 nm as functions of $B_z$ measured three times. The light red/blue areas and black horizontal lines mark the positions of small kinks in $\sigma_{12}$.

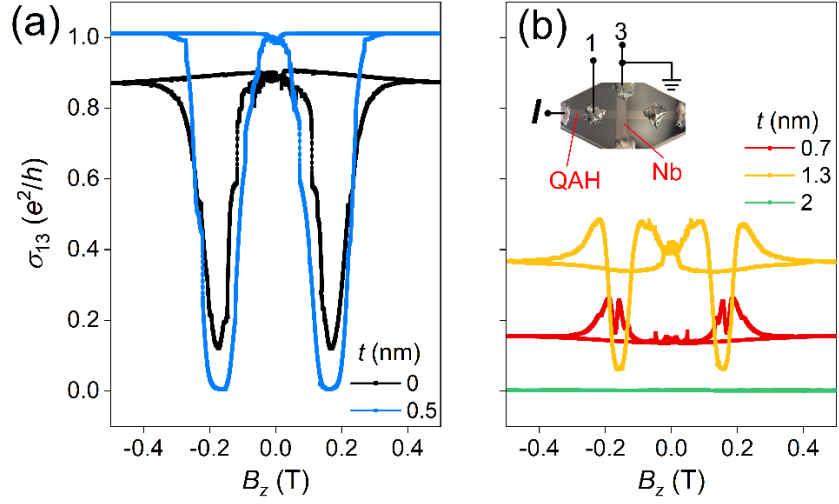

FIG. 3. (a-b) Conductance $\sigma_{13}$ of the QAH/SC/QAH junctions as functions of $B_z$ with various $t$. Small kinks/plateaus in (a) are unrepeatable and may relate to random domain fluctuations. Inset in (b) shows the measurement configuration.

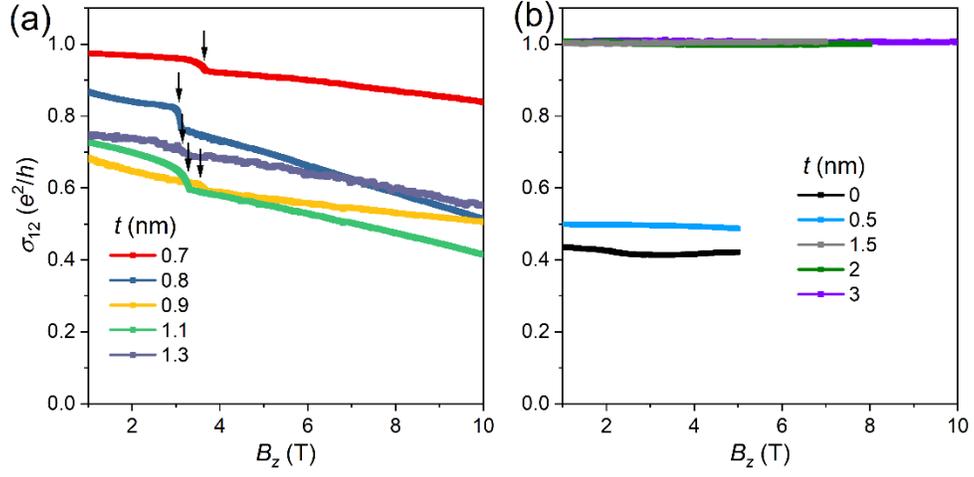

FIG. 4. (a-b) Conductances $\sigma_{12}$ of the QAH/SC/QAH junctions with various $t$ as functions of high-$B_z$. The arrows in (a) mark the positions of prominent decreases in $\sigma_{12}$.